\documentclass[article]{JHEP3}
\usepackage{amsmath,amssymb}
\usepackage{cite}
\usepackage{float}
\usepackage{comment}
\usepackage{subcaption}
\usepackage{amsfonts}
\usepackage[utf8]{inputenc}
\usepackage{graphicx}

\allowdisplaybreaks

\newcommand{\be}{\begin{equation}}
\newcommand{\ee}{\end{equation}}
\newcommand{\bea}{\begin{eqnarray}}
\newcommand{\eea}{\end{eqnarray}}
\newcommand{\bean}{\begin{eqnarray*}}
\newcommand{\eean}{\end{eqnarray*}}

\def\beq{\begin{equation}}

\def\eeq{\end{equation}}

\def\Re{\mathop{\rm Re}}
\def\Im{\mathop{\rm Im}}
\relax

\title{Eikonal quasinormal modes, greybody factors and shadow of charged accelerating black holes}

\author{Filipe Moura$^a$ and Francisco Silva$^b$
\\
\\
$^{a}$
Departamento de Matem\'atica, Escola de Tecnologias e Arquitetura, \\ ISCTE - Instituto Universit\'ario de Lisboa \\ and Instituto de Telecomunica\c c\~oes,
\\Av. das For\c cas Armadas, 1649-026 Lisboa, Portugal\\
\email{fmoura@lx.it.pt}
\\
\\
$^{b}$
Instituto de Telecomunica\c c\~oes and Departamento de F\'\i sica,\\ Faculdade de Ci\^encias da Universidade de Lisboa,\\
Campo Grande, 1749-016 Lisboa, Portugal\\
\email{fc58297@alunos.ciencias.ulisboa.pt}
}

\abstract{We show that the quasinormal modes, in the eikonal limit, for accelerating (non-rotating) black holes, are related to the angular velocity of the circular null geodesics and to the corresponding Lyapunov exponent, in a similar, but generalized way as the ones for spherically symmetric black holes are. We compute those quasinormal modes and greybody factors for neutral and charged accelerating black holes by considering massless test scalar fields. Later we show that the obtained results are universal for perturbations of any spin. We also determine the radius of the shadow cast by these black holes. Our results for charged accelerating black holes simplify to the ones of the Reissner-Nordstr\"om black hole simply by setting acceleration to zero.
}

\begin{document}



\vfill

\eject

\section{Introduction}
\noindent

According to the Kerr hypothesis, in general relativity any isolated astrophysical (uncharged) black hole is described by the Kerr metric, and therefore completely defined by just two parameters: the mass and angular momentum. However, Einstein's vacuum field equation admits more general black hole solutions than the Kerr solution, which are applicable to nonisolated objects.

Recent observational advances motivate consideration of elaborated astrophysical formation scenarios, in which black holes are formed or reside in dense environments like globular clusters, or in the vicinity of other massive bodies. These black holes are not isolated and can thus experience acceleration. Furthermore, the emission of gravitational waves by black hole binaries in the inspiral phase tends to have a preferred direction, which results in the black hole remnant having a recoil acceleration after the merger - a process called the black hole superkick \cite{Bruegmann:2007bri}.

It is therefore of interest to understand accelerating black holes and their radiation, as well as the interaction with their astrophysical
environment. From a theoretical point of view one must incorporate acceleration into exact vacuum solutions of Einstein’s field equation. This is achieved through the C-metric \cite{Weyl:1917gp}, describing an accelerating black hole.

The presence of such acceleration will generically perturb geodesic motion near the black hole, changing lensing time delays and shifting the optimal viewing inclination for shadows. Furthermore, acceleration will modify observable signatures, introducing measurable corrections to the spectra of quasinormal modes (QNMs) and greybody factors. Future observations could make it possible to distinguish Kerr black holes from accelerating black holes by their lensing features. For this reason, it is of relevance to theoretically study these properties of accelerating black holes.

The first numerical study of quasinormal modes corresponding to scalar fields for accelerating black holes has been performed in \cite{Destounis:2020pjk}, where it was shown that the spectrum consists of three distinct families: the photon-surface modes, the accelerating modes and (for black holes with charge) the near-extremal modes. Similar results were obtained through a mapping between the master equation obeyed by quasinormal modes and a quantum Seiberg-Witten curve \cite{Lei:2023mqx}: the QNM spectrum is obtained numerically through a quantization condition derived from the Nekrasov-Shatashvili free energy. Quasinormal modes of Kerr-Newman-anti-de Sitter slowly accelerating black holes were studied numerically using the method of isomonodromic deformations in \cite{BarraganAmado:2023wxt}. These studies have been extended to the spinning C-metric in \cite{Chen:2024rov}, where quasinormal modes corresponding to gravitational perturbations have been computed numerically using the continued fraction and shooting methods.

There are some limiting cases where analytical formulas for quasinormal modes and greybody factors can be obtained. One of those is the eikonal limit, corresponding to a large real part of the emitted/scattered radiation. This is the approximation corresponding to geometrical optics, where the WKB method should be applied. By taking this limit, we compute the quasinormal modes and greybody factors for accelerating black holes. Afterwards, we also determine the radius of the shadow cast by them.

The article is organized as follows. In section \ref{C} we review the C-metric - both for uncharged and charged black holes. We also consider the master equations for scalar test fields, in order to obtain the quasinormal modes and greybody factors. We take the eikonal limit of the respective potential and show that in this limit the quasinormal modes and greybody factors can be obtained for the C-metric using the WKB approximation. This is then used to generalize a result valid for spherically symmetric metrics, namely: the real part of the quasinormal frequencies is proportional to the angular velocity of the unstable circular null geodesics and the imaginary part of those frequencies is proportional to the corresponding Lyapunov exponent. In section \ref{eik} we compute these quantities, together with the radius of the shadow, for uncharged black holes. In section \ref{4} we extend these results in the presence of black hole charge. In the appendix \ref{AppendixA} we show that the results for eikonal quasinormal modes and greybody factors, obtained for scalar test fields, are also valid for perturbations of any spin.

\section{Accelerating black holes and their quasinormal modes}
\label{C}
\subsection{The C-metric}

\noindent

An accelerating black hole can be described by the C-metric, originally found in \cite{Weyl:1917gp} and given in spherical-type coordinates by
\begin{equation}
    ds^2=\frac{1}{(1-ar\cos{\theta})^2}\left(-f(r)dt^2+\frac{dr^2}{f(r)}+\frac{r^2d\theta^2}{P(\theta)}+P(\theta)r^2\sin^2{(\theta)}d\varphi^2\right),
    \label{Metric}
\end{equation}
where
\begin{equation}
    f(r)=\left(1-\frac{2M}{r}+\frac{Q^2}{r^2}\right)\left(1- a^2r^2\right) \quad \text{and} \quad P(\theta)= 1-2aM\cos{\theta}+a^2Q^2\cos^2{\theta}.
    \label{fp}
\end{equation}
The parameter $a$ represents the acceleration of the black hole, while $Q$ and $M$ represent the black hole's charge and mass parameters respectively. The transformation $a \rightarrow -a$ is equivalent to the isometry $\theta \rightarrow \pi-\theta.$ We can therefore take $a$ to be positive without loss of generality. The limit case $a=0$ reduces to the Reissner-Nordstr\"om solution.

In the metric \eqref{Metric} one can find conical singularities at the poles, i.e. for $\theta =0, \, \pi.$ One of these singularities (not both) can be removed by choosing an adequate deficit or excess angle in the variable $\varphi$ (corresponding to those values of $\theta$). If one chooses to remove the singularity at $\theta =0,$ the excess angle at $\theta =\pi$ can be interpreted as a strut pushing the black hole. If one rather chooses to remove the singularity at $\theta =\pi,$ the deficit angle at $\theta =0$ can be interpreted as a cosmic string pulling the black hole. Both configurations act as sources of acceleration for the black hole. For a detailed discussion see \cite{Destounis:2020pjk, Kinnersley:1970zw}.

The maximal analytical extension of the C-metric actually describes two causally disconnected black holes accelerating in opposite directions; however, because they are causally disconnected, we can consider only one of them. A more complete study of this metric can be found in \cite{Griffiths:2006tk}.


The metric admits as horizons
\bea
    r_\pm =M\pm\sqrt{M^2-Q^2}, \label{rpm} \\
    r_{a}=1/a. \label{rh}
\eea
Physically we expect small magnitudes for the black hole acceleration $a$, which means the acceleration horizon \eqref{rh} should be very large, namely $r_{a}\gg r_+.$ We place our observer with coordinate $r$ in the interval $r_+<r<r_{a}$.

We can express the metric (\ref{Metric}) in a more familiar way by performing a conformal transformation $g_{\mu \nu}\rightarrow \tilde g_{\mu \nu}=\Sigma^2g_{\mu \nu}$, with $\Sigma=(1-ar\cos{\theta})$:
\begin{equation}
d\Tilde{s}^2=\Sigma^2ds^2=-f(r)dt^2+\frac{dr^2}{f(r)}+\frac{r^2d\theta^2}{P(\theta)}+P(\theta)r^2\sin^2{(\theta)}d\varphi^2.
\label{confmetric}
\end{equation}
Notice that the presence of $P(\theta)\neq1$ means the spacetime is not spherically symmetric.


\subsection{Scalar test field and master equations}
\label{wave}
\noindent

For simplicity, in order to compute the quasinormal modes and greybody factor we start by considering a massless and chargeless test scalar field $\psi.$ In appendix \ref{AppendixA}, we argue that in the eikonal limit the results are the same for other massless perturbations of different spin.

In the background of the black hole, this field satisfies the Klein Gordon equation
\begin{equation}
    \Box_g\psi=\frac{1}{\sqrt{-g}}\partial_\mu\left( \sqrt{-g}g^{\mu \nu}\partial_\nu \psi\right)=0,
    \label{klein}
\end{equation}
where $g=\det(g_{\mu \nu})$.
In the background of the metric (\ref{confmetric}), $\psi$ can be factorized as
\begin{equation}
    \psi=e^{-i\omega t}e^{im\varphi}\frac{\phi(r)}{r}\chi(\theta), \label{factpsi}
\end{equation}
in such a way that the radial and angular equations of motion become separable, reading \cite{Destounis:2020pjk}
\begin{equation}
    \frac{d^2\phi(r)}{dx^2}+(\omega^2-V_r)\phi(r)=0, \quad     \frac{d^2\chi(\theta)}{dz^2}-(m^2-V_{\theta})\chi(\theta)=0,
    \label{waveeq}
\end{equation}
with $x$ and $z$ given by
\begin{equation}
    dx=\frac{dr}{f(r)},\quad dz=\frac{d\theta}{P(\theta)\sin{\theta}}.
\end{equation}
The potentials above are given by
\begin{equation}
    V_r=f(r)\left(\frac{\lambda}{r^2}-\frac{f(r)}{3r^2}+\frac{f'(r)}{3r}-\frac{f''(r)}{6}\right)
    \label{Vr}
\end{equation}
and
\begin{equation}
    V_\theta=P(\theta)\left(\lambda \sin^2{\theta}-\frac{P(\theta)\sin^2{\theta}}{3}+\frac{\sin{\theta\cos{\theta}P'(\theta)}}{2}+\frac{\sin^2{\theta}P''(\theta)}{6}\right).
    \label{Vtheta}
\end{equation}
where $\lambda$ is the separation constant of the two equations (\ref{waveeq}), $m$ is the angular azimuthal quantum number and $\omega$ is the time
frequency of the wave where our main focus resides.

Explicitly, replacing  $f(r)$ and $P(\theta)$ from (\ref{fp}), $V_r$ and $V_\theta$ simplify to
\bea
    V_r&=&f(r)\left(\frac{\lambda-1/3+a^2Q^2/3}{r^2}+\frac{2M}{r^3}-\frac{2Q^2}{r^4} \right),
    \label{expandvr} \\
    V_\theta&=&P(\theta)\left(\left(\lambda-1/3-\frac{5 a^2 Q^2}{3}\right)\sin^2\theta +2aM\cos{\theta}\sin^2{\theta} +2a^2Q^2\sin^4\theta\right).
    \label{expandvt}
\eea

Before solving these equations one needs to introduce boundary conditions. For QNMs the appropriate conditions are
\begin{equation}
    \phi(r) \sim \begin{cases}
        e^{-i\omega x},\quad x\rightarrow -\infty \; (r\rightarrow r_+)\\
        e^{+i\omega x}, \quad x\rightarrow +\infty \;(r\rightarrow1/a)
    \end{cases}.
    \label{conditions}
\end{equation}
The physical meaning of these is that nothing comes out of the black hole's horizon (we have a purely incoming wave) and nothing can come from outside the acceleration horizon (we have a purely outgoing wave).

For the angular part we choose boundary conditions in order for $\chi(\theta)$ to be regular at the poles:
\begin{equation}
    \chi(\theta) \sim \begin{cases} e^{+mz}, \quad z\rightarrow -\infty \; (\theta\rightarrow 0) \\
    e^{-mz}, \quad  z\rightarrow +\infty \;(\theta\rightarrow\pi)
    \end{cases}.
    \label{chiconditions}
\end{equation}

From the discussion following \eqref{fp}, the coordinate $\varphi$ has periodicity $2\pi C$, where $C$ is fixed by removing one of the conical singularities. In what follows we adopt the choice $C=1/P(\pi)$, corresponding to the physically relevant configuration of a cosmic string pulling the black hole. Requiring the function $e^{im\varphi}$ to be periodic leads to $e^{2\pi i m C}=1$, which would imply a redefinition of $m$ depending on the black hole parameters as $m = \hat m /C=P(\pi) \hat m=(1+2M a+ a^2Q^2) \hat m,$ $\hat m$ being an integer. Alternatively, and more simply, we prefer to redefine the coordinate $\varphi$ as $\hat \varphi = \varphi/C=P(\pi) \varphi,$ with $0\leq \hat \varphi < 2 \pi.$ With this redefinition, $m$ remains an integer.

The boundary conditions \eqref{conditions} considered make the problem non hermitian. This means that the eigenvalues $\omega^2$ of the equation we intend to solve \eqref{waveeq} no longer need to be real. In fact, it turns out that generally $\omega$ is a complex number. Physically one may say that conditions \eqref{conditions} define the black hole as a dissipative system, hence the solution of the wave equation will either exhibit a damping or exponential growth with time. This can only be achieved by complex QNM frequencies and the sign of its imaginary part will determine whether oscillations are stable (damped) or unstable (exponential growth).

Since the physical quantities of interest are $\lambda$-dependent, one must determine the values this separation constant is allowed to take. When $a=0$, we have spherical symmetry and the solution to the angular part of \eqref{waveeq} is just given by the spherical harmonics, labeled by a quantum angular momentum number $\ell$. It is well known that in this case we have $\lambda= \ell(\ell+1)+1/3$, and we expect that to remain valid in the limit $a\rightarrow0$. However, for the general $a\neq0$ case, it is difficult to obtain a complete solution for $\lambda$. There are some limit results known in the literature: in \cite{Kofron:2015gli,Kofron:2016dyk} one can find the solution with $m=0$. In \cite{BarraganAmado:2023wxt} one can find a solution with general $m\neq 0$ that is perturbative in $a$, which we write here explicitly:
\bea
\lambda&=&\ell(\ell+1) +\frac{1}{3}+\left[ \frac{1-\ell(\ell+1)(3\ell^2+3\ell-1)-m^2(15\ell^2+15\ell-11)}{2(2\ell-1)(2\ell+3)}r_+^2\left(1+\frac{Q^4}{r_+^4}\right) \right. \nonumber \\&-& \left. \frac{\ell(\ell+1)(3\ell^2+3\ell-2)+3m^2(9\ell^2+9\ell-7)}{3(2\ell-1)(2\ell+3)}Q^2\right] a^2+O(a^3).
\label{lambda}
\eea

In the eikonal limit that we consider for our calculations, we take the approximation of very large angular momentum number
$\ell\rightarrow \infty$. But here one must be careful: since $m$ varies from $-\ell$ to $\ell$, $m-$dependent terms can also contribute and should be considered in the eikonal limit.

Since the potentials depend on $\ell$ and $m$ only through $\lambda$, we should consider the asymptotic behavior of $\lambda$ when $\ell \rightarrow \infty$. From \eqref{lambda}, and considering that $r_+^2+\frac{Q^4}{r_+^2}= 4 M^2 - 2 Q^2,$ we get
\be
\lambda \sim \ell(\ell+1)\left[1-\frac{3M^2}{2} a^2+\frac{Q^2}{2} a^2+m^2\left(-\frac{15}{2} M^2a^2+\frac{3}{2} Q^2a^2\right)\right]+O(a^3). \label{lambda2}
\ee
The equation above can be written as
\begin{equation}
    \sqrt{\lambda} \sim \left(\ell+\frac{1}{2} \right)\left(1-\frac{3}{4}M^2 a^2 +\frac{1}{4} Q^2 a^2 \right)+\frac{m^2}{\ell+\frac{1}{2}}\left(-\frac{15}{4}M^2+\frac{3}{4}Q^2 \right)a^2+O(a^3).
\label{lambdasy}
\end{equation}

Both \eqref{lambda2} and \eqref{lambdasy} are asymptotic expansions in $\ell$, but also perturbative expansions in $a$. From now on, we will mostly consider $a$ to be a small perturbative parameter. 

The equations of motion (and the respective potentials) that we consider are for a massless scalar field, and so is their separation constant \eqref{lambda}. To fields of different spin correspond different potentials and different separation constants. However, in appendix \ref{AppendixA} we argue that the asymptotic expansion in the eikonal limit given by \eqref{lambdasy} is universal for the separation constants corresponding to other massless perturbations of different spin.

\subsection{Eikonal limit and WKB approximation}
\noindent

For our calculations we consider the eikonal limit, also known as the optical approximation or semiclassical limit, in which we take the approximation of very large angular momentum number $\ell\rightarrow \infty$. Since the potentials do not depend on $\ell$ explicitly but only through the separation constant $\lambda$, one must figure its behavior in the eikonal limit. From \eqref{lambdasy}, it is clear that this limit is equivalent to $\lambda \rightarrow \infty$ if we consider $a$ to be treated perturbatively. Therefore, for our practical purposes we generalize the eikonal limit as the limit of $\lambda \rightarrow \infty$. In this way we are allowed to discard any term that is not proportional to $\lambda$ in the potentials $V_r$ and $V_\theta$ given by \eqref{Vr} and \eqref{Vtheta}. This translates to
\begin{equation}
V_r(r)\approx f(r)\frac{\lambda}{r^2} =:V^{\mathrm{eik}}_r
\label{vreikonal}
\end{equation}
and
\begin{equation}
    V_\theta\approx \lambda P(\theta) \sin^2{\theta}=:V^{\mathrm{eik}}_\theta. \label{vthetaeikonal}
\end{equation}

Considering this limit, WKB theory provides a approximate solution for the quasinormal frequency $\omega$ as the eigenvalue of the radial equation \eqref{waveeq}, given by \cite{Schutz:1985km}

\begin{equation}
    \omega=\frac{\sqrt{\lambda f(r_c)}}{r_c}-i\left(n+\frac{1}{2}\right) \sqrt{-\frac{r_c^2}{2f(r_c)}\left(\frac{d^2}{dx^2}\frac{f(r)}{r^2} \right)_{r=r_c}}, \label{wkb}
\end{equation}
where $r_c$ is the unique maximum of the potential $V_r$.

\subsection{The eikonal quasinormal mode photon surface correspondence in accelerating spacetimes\label{geodesic}}
\noindent

For spherically symmetric black holes there is a known correspondence between quasinormal modes in the eikonal limit and the circular null geodesics. This correspondence can be found in \cite{Cardoso:2008bp} and it states that quasinormal frequencies in this limit are given by
\begin{equation}
    \omega=\left(\ell+\frac{1}{2} \right) \Omega_c-i\left(n+\frac{1}{2}\right)\Lambda_P \label{card}
\end{equation}
where $\Omega_c$ is the frequency of unstable circular null geodesics and $\Lambda_P$ is the Lyapunov exponent of the corresponding motion. Essential for establishing such correspondence is the fact that the equation for maximizing the potential $V^{\mathrm{eik}}_r(r)$ in the eikonal limit given by \eqref{vreikonal} is the same equation defining the circular null geodesics:
\begin{equation}
    \left(V^{\mathrm{eik}}\right)'(r_c)=0 \Leftrightarrow r_c f'(r_c)=2f(r_c),
    \label{max}
\end{equation}
where the prime denotes the derivative with respect to $r$. The radius of these circular null geodesics is therefore given by $r_c$.

For rotating black holes the correspondence must be modified: in this case, $\Omega_c$ in \eqref{card} is replaced by $\Omega_c+\frac{m}{\ell+\frac{1}{2}}\omega_{\mathrm{prec}}$, where $\omega_{\mathrm{prec}}$ is the Lense-Thirring precession frequency of the photon orbit arising due to the rotation of the black hole \cite{Yang:2012he,Li:2021zct}.

In the spherically symmetric limit $a=0$, from \eqref{lambdasy} one has asymptotically $\sqrt{\lambda} \approx \ell+\frac{1}{2}$; in this limit, identifying equation \eqref{card} with \eqref{wkb}, one gets for $\Omega_c$ and $\Lambda_P$ the expressions
\bea
&&\Omega_c=\frac{\sqrt{f(r_c)}}{r_c}, \label{O} \\
&&\Lambda_P=\sqrt{-\frac{r_c^2}{2f(r_c)}\left(\frac{d^2}{dx^2}\frac{f(r)}{r^2} \right)_{r=r_c}}. \label{L}
\eea

In this section we want to generalize the correspondence \eqref{card} and the identifications \eqref{O} and \eqref{L} for spacetimes of the form \eqref{Metric} with $P({\theta})\neq1$, like the one we consider in this article\footnote{Notice that null geodesics from \eqref{Metric} are the same as the null geodesics in \eqref{confmetric} since these are preserved under conformal transformations.}, and which were not considered in \cite{Cardoso:2008bp,Yang:2012he,Li:2021zct}.

We recall that the geodesic equations of motion for a massless test particle can be written using the Hamilton-Jacobi formalism (for a review in the context of the C-metric see \cite{Frost:2020zcy}\footnote{In article \cite{Frost:2020zcy} $Q(r)$ corresponds to our $f(r)$ and $\lambda$ corresponds to our $\eta$.}).

The metric (\ref{confmetric}) is static and axisymmetric and thus has two Killing vectors, one ($\partial_t$) associated to time translation and the other ($\partial_\varphi$) to axial symmetry. For such metrics, the Hamilton-Jacobi equation of null geodesics is separable, this allows us to obtain the following geodesic equations \cite{Frost:2020zcy}
\begin{equation}
    \frac{dt}{d\eta}=\frac{r^2E}{f(r)},
    \label{t}
\end{equation}
\begin{equation}
    \frac{d\varphi}{d\eta}=\frac{L_z}{P(\theta)\sin^2{\theta}},
    \label{phi}
\end{equation}
\begin{equation}
   \left(\frac{dr}{d\eta}\right)^2=r^4 E^2-r^2f(r)\mathcal K,
   \label{r}
\end{equation}
\begin{equation}
    \left( \frac{d\theta}{d\eta}\right)^2=P(\theta)\mathcal{K}-\frac{L_z^2}{\sin^2{\theta}},
    \label{the}
\end{equation}
where $\eta$ is the Mino parameter related to the affine parameter by
\begin{equation}
    \frac{d\eta}{ds}=\frac{\Sigma^2}{r^2},
\end{equation}
and $\Sigma=1-ar\cos\theta$ is associated to the conformal transformation performed in \eqref{confmetric}.

The integrals of motion $L_z$, $E$, and $\mathcal{K}$ arise when reducing the second-order equations of motion to first-order form. These constants represent the projected angular momentum ($L_z$), the particle energy ($E$), and the Carter constant ($\mathcal{K}$) \cite{Carter:1968rr}. Specifically, $\mathcal{K}$ generalizes the total angular momentum and emerges due to the separability of the $r$ and $\theta$ components in the Hamilton-Jacobi equation.

An equivalent way to rewrite equations $\eqref{r}$ and $\eqref{the}$ is the following:
\begin{equation}
    \frac{1}{r^4}\left(\frac{dr}{d\eta}\right)^2+\mathcal{V}_r(r)\mathcal{K}=E^2
    \label{rgeo}
\end{equation}
\begin{equation}
    \sin^2{\theta}\left(\frac{d\theta}{d\eta}\right)^2-\mathcal{V}_\theta(\theta)\mathcal{K}=-L_z^2
\end{equation}
where $\mathcal{V}_r(r)=f(r)/r^2$ and $\mathcal{V}_\theta(\theta)=P(\theta) \sin^2{\theta}$. These potentials are thus identical to those in the eikonal limit, respectively \eqref{vreikonal} and \eqref{vthetaeikonal}, up to a multiplicative constant.

To later obtain the promised correspondence, we define the angular coordinate velocity as
\begin{equation}
    \Omega_c=\frac{\sqrt{\dot\theta^2/P(\theta)+\dot \varphi^2P(\theta)\sin^2\theta}}{\dot t}.
    \label{angularvelocity}
\end{equation}
By the chain rule $\frac{d}{d\tau}=\frac{d\eta}{d\tau}\frac{d}{d\eta}$, \eqref{angularvelocity} can be written as
\begin{equation}
    \Omega_c=\frac{\sqrt{\left(\frac{d\theta}{d\eta}\right)^2/P(\theta)+\left(\frac{d\varphi}{d\eta}\right)^2 P(\theta)\sin^2{\theta}}}{\frac{dt}{d\eta}}.
    \label{frequency}
\end{equation}
To simplify this equation we notice that, from \eqref{r}, when a geodesic has constant radial coordinate $r=r_c$ we get
\be
E^2=f(r_c)\mathcal{K}/r_c^2. \label{e2}
\ee
Then, by just substituting \eqref{t}, \eqref{phi} and \eqref{the} in \eqref{frequency} one obtains (\ref{O}):
\begin{equation}
    \Omega_c=\frac{\sqrt{\mathcal{K}}}{r_c^2E}f(r_c)=\frac{\sqrt{\mathcal{K}}}{\sqrt{f(r_c)\mathcal{K}}}\frac{f(r_c)}{r_c}=\frac{\sqrt{f(r_c)}}{r_c},
\end{equation}
which gives the angular velocity of circular null geodesics.

The identification (\ref{L}) for the Lyapunov exponent can be obtained by analyzing the instability of the radial equation of motion. We take the coordinate radius $r_c$ and perturb it slightly with $\delta r\ll r_c$, such that $r=r_c+\delta r$. We can find out the equation of motion for $\delta r$ by plugging this ansatz in \eqref{rgeo} and considering terms only up to second order in $\delta r$. This leads to
\begin{equation}
    \frac{1}{r_c^4}\left( \frac{d \delta r}{d\eta} \right)^2=-\frac{\mathcal{K}\mathcal{V}_r''(r_c)}{2}(\delta r)^2. \label{perturbedgeodesic}
\end{equation}
This equation admits a solution of the exponential form
\begin{equation}
    \delta r =Ae^{\Lambda_P t},
    \label{exponential}
\end{equation}
where $A$ is an integration constant and $\Lambda_P$ is defined as the Lyapunov exponent which gives the instability time scale the geodesic in consideration. Using \eqref{exponential} in \eqref{perturbedgeodesic} one finds that
\begin{equation}
    \Lambda_P^2=-\frac{r_c^4}{2\left(\frac{dt}{d\eta}\right)^2}\mathcal{K}\mathcal{V}_r''(r_c).
    \label{lambdaP}
\end{equation}
By further simplifying this with the relations \eqref{t} and \eqref{e2}, considering only the positive root \eqref{lambdaP} we conclude that
\begin{equation}
    \Lambda_P=\sqrt{\frac{-r_c^2f(r_c)}{2}\left( \frac{d^2}{dr^2}\frac{f(r)}{r^2}\right)_{r=r_c}}=\sqrt{\frac{-r_c^2}{2f(r_c)}\left( \frac{d^2}{dx^2}\frac{f(r)}{r^2}\right)_{r=r_c}},
    \label{Lyapunov Exponent}
\end{equation}
where in the last equality the coordinate $r$ was switched to the tortoise coordinate $x$ and we used the fact that $\mathcal{V}'(r_c)=0$.

To sum up, we have discovered that the Lyapunov exponent of null geodesics given by \eqref{Lyapunov Exponent} coincides with the factor present in the imaginary part of the quasinormal mode \eqref{wkb}. Analogously, we have also found that the angular velocity of the unstable circular null geodesics coincides with the factor present in the real part of the quasinormal mode \eqref{wkb}. Therefore, we infer that the generalized eikonal quasinormal mode photon surface correspondence in accelerating spacetimes should be given by
\begin{equation}
    \omega =\sqrt{\lambda}\Omega_c-i\left(n+\frac{1}{2}\right)\Lambda_P.
\end{equation}

And equivalent path to derive this correspondence would be to look at the Penrose limit. The metric (\ref{confmetric}) is of the form of a general static metric considered in equation (3.1) of reference \cite{Giataganas:2024hil}, where the Penrose limit of such metrics is discussed; therefore, we can directly apply the results obtained in such article in order to obtain the identifications (\ref{O}) and (\ref{L}) for the angular velocity and Lyapunov exponent of the circular null geodesics, in the Penrose limit of the metric (\ref{confmetric}).

\section{Eikonal limit for an uncharged accelerating black hole\label{eik}}
\noindent

We now proceed to computing the quasinormal frequencies, black hole shadow and greybody factor. In this section, for simplicity, we will consider the black hole charge in \eqref{fp} to be $Q=0$. However, the procedures we take are conceptually straightforward to generalize for a nonzero charge; the problem is just more complicated algebraically. That generalization will be conducted in section \ref{4}. The potentials $V_r$ and $V_\theta$ in the eikonal limit, given by \eqref{vreikonal} and \eqref{vthetaeikonal}, translate to
\begin{equation}
V^{\mathrm{eik}}_r =\frac{\lambda}{r^2} \left(1-\frac{2M}{r}\right)\left( 1- a^2r^2\right)
\label{vreikonal2}
\end{equation}
and
\begin{equation}
V^{\mathrm{eik}}_\theta=\lambda\sin^2{\theta}(1-2aM\cos{\theta}). \label{vthetaeikonal2}
\end{equation}
We need to find the maximum of $V^{\mathrm{eik}}_r$. This requires us to solve \eqref{max}, which is equivalent to
\begin{equation}
r_c+M(-3+a^2~r_c^2)=0.
   \label{rc}
\end{equation}
Solving \eqref{rc}, we obtain
\begin{equation}
    r_c=
    \frac{6M}{1+\sqrt{1+12a^2M^2}}= 3M-9M^3a^2+54M^5 a^4+\mathcal{O}(a^6).
    \label{rcrit}
\end{equation}
Considering the limit $a\rightarrow0,$ we obtain the result corresponding to the Schwarzschild solution.

\subsection{Calculation of the quasinormal frequencies}
\noindent

Let us now proceed to calculate the quasinormal frequencies $\omega$.

Taking $r_c$ given by \eqref{rcrit} in \eqref{wkb} and \eqref{L}, we obtain the real and imaginary parts of $\omega$ as
\bea
&&\Re(\omega)=
\sqrt{\lambda}\frac{\sqrt{1+\sqrt{1+12a^2M^2}+12a^2M^2(-3+\sqrt{1+12a^2M^2})}}{3\sqrt{6}M}, \label{Rew} \\
&&\Lambda_P=\frac{\sqrt{1+\sqrt{1+12a^2M^2}+12a^2M^2(2+12a^2M^2-3\sqrt{1+12a^2M^2})}}{3\sqrt{6}M};
\label{im}
\eea
taking $\sqrt{\lambda}$ given by \eqref{lambdasy} (with $Q=0$) and expanding in powers of $a$ we get
\bea
&&\Re(\omega)=\left(\ell+\frac{1}{2}\right)\left(\frac{1}{3\sqrt{3}M}-\frac{7}{4\sqrt{3}}Ma^2 \right)-\frac{m^2}{\ell+\frac{1}{2}}\frac{5Ma^2}{4\sqrt{3}} +\mathcal{O}(a^3), \label{rea}\\
&&\Lambda_P= \frac{1}{3\sqrt{3}M}-\frac{1}{2\sqrt{3}}Ma^2-\frac{63}{8\sqrt{3}}M^3a^4+\mathcal{O}(a^6).\label{lambda_Pa}
\eea

Both for $\Re(\omega)$ and $\Lambda_P$ we observe the first order terms correspond to the QNM solution of the Schwarzschild black hole, while the higher order terms introduces corrections in $a$. The second order correction decreases the real part frequency as well as $\Lambda_P$, which therefore increases $\Im(\omega)$, since $\Im(\omega)\propto -\Lambda_P$. This leads to the interpretation that acceleration decreases both oscillation frequencies and the damping rate of quasinormal modes, a phenomenon analogous to the Schwarzchild de Sitter black hole.

However, not everything is similar, as the lack of spherical symmetry of the spacetime associated to the acceleration creates a splitting effect caused by the $m^2$ dependence in $\Re(\omega)$ which can be seen from \eqref{rea}. This splitting effect therefore not only provides a way to distinguish accelerating black holes from black holes living in de Sitter space, but also offers a new potential observational signature of the acceleration of the black hole.

\subsection{Shadow of the black hole}
\label{shadow}
\noindent

We now proceed to determining the radius of the shadow cast by this black hole and relating it to the quasinormal frequencies, according to the eikonal correspondence \cite{Chen:2022nlw}.

In order to define the shadow of a black hole, we suppose that the universe is filled with light sources in all directions except for the region directly between the observer and the black hole.
The radius $r_c$ of the photon surface corresponds to an unstable circular orbit for light; therefore, light rays that start slightly inside this orbit will inevitably be pulled into the black hole, while those slightly outside will scatter away to infinity. As a result, the dark region we observe in images of black holes is not the event horizon itself, but rather the projection of the photon surface, which determines the apparent size of the black hole's shadow.

The angular shadow radius of an accelerating black hole can be defined as the inverse sine of the black hole shadow radius $\mathfrak{R}$ over the distance of the observer to the black hole:
\begin{equation}
    \text{angular shadow radius}=\arcsin{\left(\mathfrak{R} \frac{\sqrt{f(r_O)}}{r_O}\right)},
    \label{angularshadowradius0}
\end{equation}
where $r_O$ is radial coordinate of the distant observer. The intrinsic shadow radius, which is independent of the observer's location, is given by
\begin{equation}
\mathfrak{R}=\frac{r_c}{\sqrt{f(r_c)}},
\label{sombras}
\end{equation}
where $r_c$ is defined by \eqref{max}.
A detailed derivation of these relations can be found in \cite{Frost:2020zcy}.


Furthermore, we can notice that \eqref{sombras} hides another correspondence between eikonal QNM and observable signatures, as from \eqref{wkb} and \eqref{O} one can write \eqref{sombras} as
\begin{equation}
    \mathfrak{R}=\frac{1}{\Omega} =\frac{\sqrt{\lambda}}{\Re(\omega)}.
    \label{shadow correspondence}
\end{equation}

A similar identification was previously made for spherically symmetric black holes in \cite{Jusufi:2019ltj} and subsequently generalized to rotating black holes in \cite{Jusufi:2020dhz,Pedrotti:2024znu}. The corresponding generalization to accelerating black holes is presented in \eqref{shadow correspondence}.

Using \eqref{rcrit}, the radius of the black hole shadow is given in our case by
\begin{equation}
\mathfrak{R}=\frac{3\sqrt{6}M}{\sqrt{1+\sqrt{1+12a^2M^2}+12a^2M^2(-3+\sqrt{1+12a^2M^2})}}=3\sqrt{3}M+\frac{81}{2\sqrt{3}}M^3a^2
+\frac{1215}{8\sqrt{3}}M^5a^4+\mathcal{O}(a^6).
    \label{shadowradiusa}
\end{equation}
Essentially, acceleration increases the shadow radius of the black hole, which can be easily seen from the $a$ perturbation expansion in \eqref{shadowradiusa}.

Further discussions on gravitational lensing features of accelerating black holes and properties of their shadows can be found in \cite{Frost:2020zcy,Grenzebach:2015oea}, including a discussion of why this shadow should be circular, even if the metric is not spherically symmetric (like in our case). Essentially the reason is that the radius of the black hole shadow depends on the integrals of motion $E$ and $\mathcal{K}$, but not on $L_z.$

%
%

\subsection{Greybody factor}
\label{greybodyfactor}
\noindent

Considering that the radiation emitted by the black hole is composed of the massless scalar test particles we have been considering, we shall now compute its corresponding greybody factor. We will assume that at the horizon ($r\rightarrow r_+$) radiation is reflected and far away ($r\rightarrow 1/a$) some of the radiation is transmitted. Mathematically this translates to considering the radial field equation \eqref{waveeq}, but with the boundary conditions \cite{Parikh:1999mf}
\begin{equation}
        \phi = \begin{cases}
        e^{i\omega x}+R e^{-i \omega x}, \quad x\rightarrow -\infty \quad(r\rightarrow r_+) \\
            Te^{i\omega x}, \quad x\rightarrow +\infty \quad (r\rightarrow1/a)
        \end{cases},
\end{equation}
where we introduce $T$ as the transmission coefficient and $R$ as the reflection coefficient. In order for energy to be conserved it is required that
\begin{equation}
    |T|^2+|R|^2=1.
\end{equation}

In the eikonal limit $\lambda\rightarrow\infty$ one can apply the WKB method \cite{Konoplya:2019hlu,Konoplya:2023moy} to obtain
\begin{equation}
    \Gamma(\omega)=\left|T\right|^2=\frac{1}{1+e^{2\pi i K}}
    \label{T}
    \end{equation}
where $K$ is defined as
\begin{equation}
    K= -i\frac{V^{\mathrm{eik}}(r_c)-\omega^2}{\sqrt{-2 \left(\frac{d^2V^{\mathrm{eik}}(r)}{dx^2}\right)_{r=r_c}}}=-i\frac{\omega^2-\Re(\omega_{n=0})^2}{4\Re{(\omega_{n=0})}\Im{(\omega_{n=0})}}.
    \label{KK}
\end{equation}
Here $\omega_{n=0}$ refers to the quasinormal with overtone number $n=0$ and the last equality shows us the correspondence between greybody factors and QNMs \cite{Konoplya:2024lir,Oshita:2024fzf,Pedrotti:2025idg}.

Therefore, in order to obtain the greybody factor one must compute $K.$ Using $\left(\frac{d^2V^{\mathrm{eik}}(r)}{dx^2}\right)_{r=r_c}=f^2(r_c)\frac{d^2V^{\mathrm{eik}}(r)}{dr^2}\big|_{r=r_c}$, equation \eqref{KK} can be recast into the form
\begin{equation}
    K=-i\left(\frac{\lambda}{r_c^2\sqrt{-2 \frac{d^2V^{\mathrm{eik}}}{dr^2}|_{r=r_c}}}-\frac{\omega^2}{f(r_c)\sqrt{-2 \frac{d^2V^{\mathrm{eik}}}{dr^2}|_{r=r_c}}}\right)=-i\left(\lambda-\frac{r_c^2}{f(r_c)}\omega^2\right)\frac{1}{r_c^2\sqrt{-2 \frac{d^2V^{\mathrm{eik}}}{dr^2}|_{r=r_c}}}.
    \label{k}
\end{equation}

Therefore, by plugging our solution for $r_c$ given by \eqref{rcrit}, we get
\begin{equation}
K=\frac{-i}{(1+12a^2M^2)^{1/4}} \left( \frac{\sqrt{\lambda}}{2}-\frac{27M^2\omega^2}{\sqrt{\lambda}(2-\sqrt{1+12a^2M^2})(1-12a^2M^2+\sqrt{1+12a^2M^2}}\right);
\end{equation}
taking $\sqrt{\lambda}$ given by \eqref{lambdasy} (with $Q=0$) and expanding in powers of $a$ we get finally
\begin{equation}
    K= -i\left[\left( \ell+\frac{1}{2}\right)\left(\frac{1}{2}-\frac{15}{8}M^2a^2 \right)-\frac{m^2}{\ell+\frac{1}{2}}\frac{15M^2a^2}{8}-\frac{\omega^2M^2}{\ell+\frac{1}{2}}\left(\frac{27}{2}+\frac{729}{8}M^2a^2 \right)-\frac{m^2\omega^2}{(\ell+\frac{1}{2})^2}\frac{405M^4a^2}{8}\right]+\mathcal{O}(a^3).
\end{equation}

To analyze the effect of acceleration on the greybody factor, it is useful to look at the correspondence between QNMs and greybody factors given by \eqref{KK}. We have seen that the real part of the QNM frequency $\Re{(\omega)}$ decreases with increasing acceleration and therefore \eqref{KK} allows us to infer qualitatively what happens to the greybody factor. The usual behavior of $\Gamma(\omega)$ defined in \eqref{T} for appropriate $K$ given by \eqref{KK} is that of a function that starts relatively close to zero as $\omega\to 0$, grows sharply in the neighborhood of $\Re{(\omega_{n=0})}$ and then approaches the value $1$ asymptotically as $\omega \to +\infty$. Therefore decreasing $\Re(\omega_{n=0})$ will essentially shift this transition point, allowing for a bigger range where $\Gamma(\omega)\sim 1$, consequently meaning that the black hole absorbs less Hawking radiation leading to its thermal spectrum to look closer to that of a black body. One can see this effect visually in figure \ref{greybodyfactor}.
\begin{figure}[h!]
    \centering
    \includegraphics[width=0.7\linewidth]{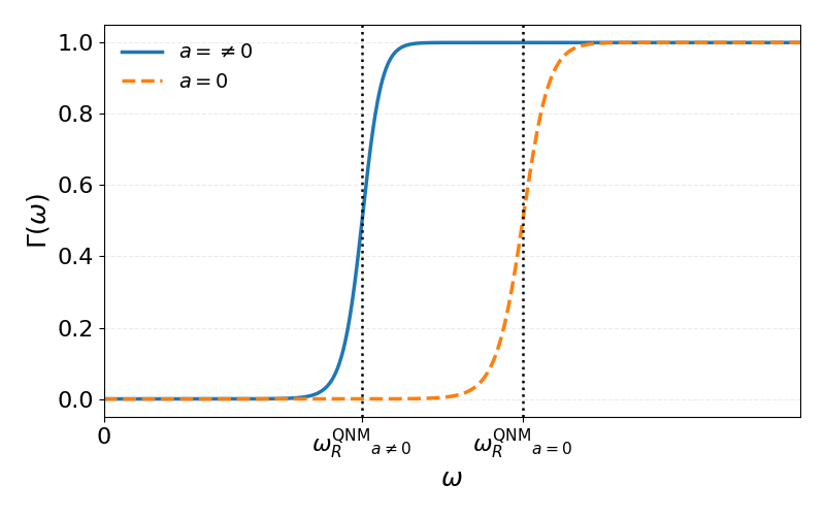}
    \caption{Greybody factor in the eikonal limit of accelerating vs non accelerating black holes. The greybody factor $\Gamma(\omega)$ is plotted as a function of the frequency $\omega$. The full line shows $\Gamma(\omega)$ of a black hole that has acceleration $a\neq0$, while the dashed line shows $\Gamma(\omega)$ for a nonaccelerating black hole. The vertical lines represent the values where $\omega$ takes the value of the real part of the QNM frequency, denoted by $\omega_{\mathrm{R}}^{\mathrm{QNM}}$, both when $a\neq0$ and $a=0$.}
    \label{greybodyfactor}
\end{figure}

\section{Eikonal limit for an accelerating black hole with charge}
\label{4}
\noindent

In this section, we will apply the same techniques in order to compute the quantities quantities of interest for a charged black hole: we take the metric \eqref{confmetric} with $f(r)$ given by \eqref{fp}, where $Q\neq0$. Conceptually everything is the same; however, the problem gets more difficult.

First, we should find the maximum of $V_r$ in the eikonal limit given by (\ref{vreikonal}). Including charge will introduce a new turning point in this potential. However we can show that the black hole will continue having only one maximum as long as we are outside of the event horizon and the condition $|Q|<M$ is verified, preserving thus the validity of the WKB method.

To see why, we notice that the potential can be written as
\begin{equation}
    V^{\mathrm{eik}}_r(r)=\lambda \frac{f(r)}{r^2}=\lambda\frac{(r-r_+)(r-r_-)(r_a-r)(r+r_{a})}{r^4r_a^2}.
    \label{eikonal potential}
\end{equation}

Considering only $r>0$, it is obvious that the $V^{\mathrm{eik}}_r(r)$ changes sign at the horizons $r_\pm$ and $r_a$ given by \eqref{rpm} and \eqref{rh}. As $r\rightarrow0$ the potential diverges to $+\infty$; at $r_-<r< r_+$ it becomes negative and attains a local minimum. When $r_+<r<r_a$ it becomes positive and goes to zero as $r\rightarrow r_a$, so there has to be local maximum in this region and, as we have seen, it must be the only one of $V_r^{\mathrm{eik}}(r)$.

As in section \ref{eik}, we will consider a test scalar field in the eikonal limit and compute its QNM spectra together with its greybody factor. In appendix \ref{AppendixA} we show that, like for the uncharged black hole we have considered previously, the results to be obtained in this section are generalizable for perturbations of arbitrary spin.

\subsection{Determining the maximum of the potential}
\noindent

The equation that gives the maximum of the eikonal potential is \eqref{max} with $f(r)$ given by \eqref{fp}. This translates to
\begin{equation}
    2Q^2-3Mr_c+(1-a^2Q^2)r_c^2+a^2Mr_c^3=0.
    \label{cube}
\end{equation}
This is a cubic polynomial equation, which in general is not easy to solve. We start by introducing the new variable $y$ as $r=y-(-a^2Q^2+1)/3Ma^2,$ getting the equivalent expression
\begin{equation}
    y^3-\frac{1+a^4Q^4+a^2(9M^2-2Q^2)}{3a^4M^2}y+\frac{2+a^2(-6Q^2+6a^2Q^4-2a^4Q^6+ 27M^2(1+a^2Q^2))}{27a^6M^3}=0. \label{pcube}
\end{equation}

This cubic equation is in the standard form $y^3+py+q=0.$ Analyzing its discriminant
\begin{equation}
D=-(4p^3+27q^2),
\end{equation}
we conclude that it is positive for all the relevant values of the black hole's parameters.
This means that the polynomial has three positive roots, given by

\begin{equation}
    y=r_c+\frac{-a^2Q^2+1}{3a^2M}=\sqrt{\frac{-4p}{3}}\cos{\left(\left(\arccos{\left(\frac{3q}{2p}\sqrt{\frac{-3}{p}}\right)}-2k\pi)\right)/3\right)}, \,\, k=0,1,2.
\end{equation}
One can perform a consistency check, and verify that indeed only one of the candidate solutions obtained for $r_c$ is greater than the horizon radius. Also, in the limit $Q\rightarrow 0$ we have to get the solution \eqref{rcrit} obtained in the previous section.
Both criteria only hold choosing $k=0.$
Therefore we conclude that we get the maximum of the potential $V_r^{\mathrm{eik}}(r)$ for
\bea
    r_c&=&\frac{-1+a^2Q^2}{3a^2M}+\frac{2\sqrt{1+a^4Q^4+a^2(9M^2-2Q^2)}}{3a^2M} \times \nonumber \\ &\times& \cos{\left(\frac{1}{3}\arccos{\left[\frac{-2+a^2(-27M^2(1+a^2Q^2)+2Q^2(3-3a^2Q^2+a^4Q^4))}{2(1+a^4Q^4+a^2(9M^2-2Q^2))^{3/2}}\right]}\right)}.
\eea
We can expand $r_c$ in powers of $a$, like we did previously, considering a perturbative expansion:
\be
r_c = \frac{3M+\sqrt{9M^2-8Q^2}}{2}+\frac{a^2}{2}\left(\frac{-27M^4+27M^2Q^2-4Q^4}{\sqrt{9M^2-8Q^2}}-9M^3+5MQ^2\right) +\mathcal{O}(a^4). \label{rcq}
\ee

\subsection{Quasinormal frequencies}
\noindent

Given that we found $r_c$ in \eqref{rcq} and we have a formula for the separation constant in \eqref{lambdasy}, we can then proceed with the calculation of the quantities we are interested in as functions of $M, \, Q$ and (perturbatively) $a$.

The real part of quasinormal frequencies becomes
\bea
    \Re{(\omega)}&=& \sqrt{\lambda} \frac{\sqrt{(Q^2-2Mr_c+r_c^2)(1-a^2r_c^2)}}{r_c^2} \nonumber \\
    &=&\left(\ell+\frac{1}{2} \right)\frac{\sqrt{6M^2+2M\sqrt{9M^2-8Q^2}-4Q^2}}{(3M+\sqrt{9M^2-8Q^2})^2}
    \Bigg( 2-\nonumber\\
    &-&a^2\left[\frac{(3M+\sqrt{9M^2-8Q^2})^2}{4}+\frac{3M^2-Q^2}{2} +
    \frac{m^2}{\ell+\frac{1}{2}}\frac{15M^2-3Q^2}{2} \right]\Bigg)\\ &+&\mathcal{O}(a^3)\nonumber;
    \label{rewQ}
\eea
for the imaginary part we have
\bea
\frac{\Im{(\omega)}}{n+\frac{1}{2}}&=&-\frac{\sqrt{(Q^2-2Mr_c+r_c^2)(1-a^2r_c^2)(-10Q^2+12Mr_c+3(-1+a^2Q^2)r_c^2-2a^2Mr_c^3)}}{r_c^3} \nonumber \\
&=& -\sqrt{27M^4-33M^2Q^2+8Q^4+M(9M^2-7Q^2)\sqrt{9M^2-8Q^2}} \left[\frac{4\sqrt{2}}{\left(3M+\sqrt{9M^2-8Q^2}\right)^3} \right. \nonumber \\
&-& \left. a^2\left( \frac{45M^4-75M^2Q^2+32Q^4}{\sqrt{2}(9M^2-8Q^2)(9M(5M^2-4Q^2)+(15M^2-8Q^2)\sqrt{9M^2-8Q^2})}\right) \right] +\mathcal{O}(a^4).\label{imwQ}
\eea

The results for $a=0$ represent the real and imaginary parts of the quasinormal frequencies of the Reissner-Nordstr\"om black hole. Obviously the leading order correction to $\Re{(\omega)}$ is negative. In order to check that the leading order correction to $\Im{(\omega)}$ is always positive for any $|Q|<M$, we can define the variable $y=Q^2/M^2$ in such a way that
\begin{equation}
    45M^4-75M^2Q^2+32Q^4=M^4(45-75y+32y^2).
\end{equation}
It now suffices to show that the polynomial $p(y) =32y^2-75y+45$ is positive for $0<y<1$. This can be done with a plot: $p(y)$ is an upward parabola which at the boundary takes the values $p(0)=45$ and $p(1)=2$. The absolute minimum is obtained when $\frac{dp}{dy}=0$ for $y=75/64,$ which lies outside the interval $]0,1[$. Even so, one can check that $p(75/64)>0$, which means that $p(y)$ is never negative, implying that the first order correction to $\Im{(\omega)}$ is always positive. Additionally, it is important to mention that the justification for why acceleration has such an effect should not change only because charge was introduced. Such justification was mentioned right after equation\eqref{lambda_Pa}.

\subsection{Comparison to numerical results}
\noindent

Quasinormal modes for the C-metric can be classified into various families according to unique properties that they share with each other, namely: accelerating modes (purely damped modes described by the Rindler surface gravity), near extremal modes (described by the surface gravity of the near extremal horizon of the black hole) and the photon surface modes (described by the angular frequency and instability timescale of unstable circular null geodesics). In ref. \cite{Destounis:2020pjk} one can find a numerical calculation of QNM frequencies corresponding to these different families of modes.

It is certainly interesting to compare the numerical results of \cite{Destounis:2020pjk} concerning the photon surface modes we have been considering in this work to our analytical results. The results can be summarized in table \ref{Numerical comparison}.

\begin{table}[h!]
\centering

\begin{tabular}{|c|c|c|}\hline
\multicolumn{3}{|c|}{$Q/M = 0.3$} \\
\hline
$\ell$ & $a M = 0.05$ & $a M = 0.1$ \\
\hline

0 &
\begin{tabular}{c}
$\omega_{\text{PS}}^\text{N}= 0.1112 - 0.1042\, i$ \\
$\omega_{\text{PS}}^\text{T} =0.0965-0.0963\,i$
\end{tabular}
&
\begin{tabular}{c}
$\omega_{\text{PS}}^\text{N}=  0.1079 - 0.1012\, i$ \\
$\omega_{\text{PS}}^\text{T} =0.0928-0.0951\,i$
\end{tabular}
\\ \hline

1 &
\begin{tabular}{c}
$\omega_{\text{PS}}^\text{N}=  0.2941 - 0.0976\, i$ \\
$\omega_{\text{PS}}^\text{T} =0.2895-0.0963\,i$
\end{tabular}
&
\begin{tabular}{c}

$\omega_{\text{PS}}^\text{N}=  0.2839 - 0.0956\, i$ \\

$\omega_{\text{PS}}^\text{T} =0.2784-0.0951\,i$

\end{tabular}
\\ \hline

2 &
\begin{tabular}{c}

$\omega_{\text{PS}}^\text{N}=  0.4852 - 0.0968\, i$ \\

$\omega_{\text{PS}}^\text{T} =0.4824-0.0963\,i$

\end{tabular}
&
\begin{tabular}{c}

$\omega_{\text{PS}}^\text{N}= 0.4674 - 0.0954\, i$ \\

$\omega_{\text{PS}}^\text{T} =0.4639-0.0951\,i$

\end{tabular}
\\

\hline
\hline
\multicolumn{3}{|c|}{$Q/M = 0.999$} \\
\hline
$\ell$ & $a M = 0.3$ & $a M = 0.5$ \\
\hline

0 &
\begin{tabular}{c}

$\omega_{\text{PS}}^\text{N}= 0.1117 - 0.0814\, i$ \\

$\omega_{\text{PS}}^\text{T} =0.0967-0.0778\,i$
\end{tabular}
&
\begin{tabular}{c}

$\omega_{\text{PS}}^\text{N}= 0.0731 - 0.0581\, i$ \\

$\omega_{\text{PS}}^\text{T} =0.0466-0.0548\,i$

\end{tabular}
\\ \hline

1 &
\begin{tabular}{c}

$\omega_{\text{PS}}^\text{N}= 0.3018 - 0.0782\, i$ \\

$\omega_{\text{PS}}^\text{T} =0.2900-0.0778\,i$
\end{tabular}
&
\begin{tabular}{c}

$\omega_{\text{PS}}^\text{N}= 0.1938 - 0.0553\, i$ \\

$\omega_{\text{PS}}^\text{T} =0.1397-0.0548\,i$
\end{tabular}
\\ \hline

2 &
\begin{tabular}{c}

$\omega_{\text{PS}}^\text{N}= 0.4987 - 0.0780\, i$ \\

$\omega_{\text{PS}}^\text{T} =0.4834-0.0778\,i$
\end{tabular}
&
\begin{tabular}{c}
$\omega_{\text{PS}}^\text{N}=  0.3196 - 0.0550\, i$ \\

$\omega_{\text{PS}}^\text{T} =0.2328-0.0548\,i$
\end{tabular}
\\ \hline
\end{tabular}

\caption{Fundamental ($n=0$) quasinormal modes for different values of $Q/M$ and $a M$. Values obtained from the WKB approximation are labeled by $T$ and values obtained numerically in \cite{Destounis:2020pjk} are labeled by $N$.}
\label{Numerical comparison}
\end{table}

Considering that the eikonal limit is an approximation valid for high angular momentum number $\ell\gg1$, the numerical and our eikonal modes naturally converge as $\ell$ increases. Since the modes presented in \cite{Destounis:2020pjk} correspond to $\ell=0-2$, some discrepancy is to be expected. Furthermore, in \eqref{lambdasy} we treated the term $\sqrt{\lambda}$ perturbatively in $a$; therefore for high acceleration, the real part of the quasinormal frequency will naturally deviate from the numerical result. In contrast, the imaginary part of the quasinormal frequency is independent of $\ell$, which implies greater accuracy even for small values of $\ell$. Nonetheless, we conclude that the numerical and analytical results remain in good agreement even in the small-$\ell$ regime.

\subsection{Shadow of the black hole}
\noindent

The previous discussion on the angular velocity of circular null geodesics $\Omega_c$ and on the radius of the black hole shadow $\mathfrak{R}$ remains valid in this case; therefore it is straightforward to obtain them using \eqref{rewQ} and with the aid of \eqref{sombras}. The result is
\be
\mathfrak{R} =\frac{\left(3M+\sqrt{9M^2-8Q^2} \right)^2}{2\sqrt{6M^2+2M\sqrt{9M^2-8Q^2}-4Q^2}}\left( 1+\frac{a^2}{8}\left(3M+\sqrt{9M^2-8Q^2}\right)^2\right)+\mathcal{O}(a^4).
\ee
The result for $a=0$ represents the radius of the shadow of the Reissner-Nordstr\"om black hole. Obviously the leading order correction to $\mathfrak{R}$ is always positive.

\subsection{Greybody factors}
\noindent

The computation of the greybody factors for a charged accelerating black hole is again similar to its uncharged version considered in section \ref{greybodyfactor}. From \eqref{k} we get
\be
K=-i \left(\sqrt{\lambda}-\frac{\omega^2r_c^4}{\sqrt{\lambda}(Q^2-2Mr_c+r_c^2)(1-a^2r_c^2)}\right) \frac{r_c}{2\sqrt{-10Q^2+12Mr_c+3(-1+a^2Q^2)r_c^2-2a^2Mr_c^3}}.
\ee

Plugging our solution for $r_c$ given by \eqref{rcq}, $\sqrt{\lambda}$ given by \eqref{lambdasy}, defining
\begin{equation}
\Delta \equiv \sqrt{9M^2 - 8Q^2},
\qquad
\Xi \equiv (3M^2 - 2Q^2 + M\Delta)\sqrt{9M^2 - 8Q^2 + 3M\Delta}
\end{equation}
and expanding in powers of $a$, we get

\bea
K &=& \frac{-i}{\Xi}\Bigg[\left(\ell+\frac{1}{2}\right)\Bigg(\frac{9M^3-7MQ^2+(3M^2-Q^2)\Delta}{\sqrt{2}} - \\
&-&\sqrt{2}a^2\Bigg\{ \Big(4374M^{10}-12393M^8Q^2+12798M^6Q^4-5721M^4Q^6 +982M^2Q^8-32Q^{10}+\nonumber \\&+&(1458M^9-3483M^7Q^2+2862M^5Q^4-915M^3Q^6+86MQ^8 )\Delta\Big)/(\Delta \Xi)+ \nonumber \\
&+& \frac{(3M^2-Q^2)\left(9M^3-7MQ^2+(3M^2-Q^2)\Delta\right)}{8}\Bigg\}\Bigg) \nonumber\\
&+&\frac{m^2a^2}{\ell+\frac{1}{2}}(-15M^2+3Q^2)\left(\frac{(9M^3-7MQ^2+(3M^2-Q^2)\Delta)}{4\sqrt{2}}\right) \nonumber \\
&-&\frac{\omega^2}{\ell+\frac{1}{2}}\Bigg( \frac{243M^5 - 270M^3Q^2 + 60MQ^4 +\left( 81M^4- 54M^2Q^2+ 4Q^4\right)\Delta}{\sqrt{2}}+ \nonumber\\
&+&\frac{3\sqrt{2}}{\Delta \Xi}a^2\Bigg\{\Big(78732 M^{12} - 255879 M^{10} Q^2 + 321489 M^8 Q^4 - 193590 M^6 Q^6 + 55512 M^4 Q^8 - 6264 M^2 Q^{10} \nonumber\\
&+& 128 Q^{12}+\left(26244 M^{11} - 73629 M^9 Q^2 + 77031 M^7 Q^4 - 36414 M^5 Q^6 + 7336 M^3 Q^8 - 440 M Q^{10} \right) \Delta\Big)+ \nonumber \\
&+&\frac{(243M^5 - 270M^3Q^2 + 60MQ^4 +\left( 81M^4- 54M^2Q^2+ 4Q^4\right)\Delta)}{24}(3M^2-Q^2) \Delta \Xi \Bigg\}\Bigg) \nonumber \\
&-&\frac{m^2\omega^2a^2}{(\ell+\frac{1}{2})^2}(15M^2-3Q^2)\left(\frac{243M^5 - 270M^3Q^2 + 60MQ^4 +\left( 81M^4- 54M^2Q^2+ 4Q^4\right)\Delta}{4\sqrt{2}}\right)
\Bigg]+\mathcal{O}(a^3). \nonumber
\eea

This allows us to compute $\Gamma(\omega)$ using \eqref{T}. Again, the result for $a=0$ represents the greybody factor of the Reissner-Nordstr\"om black hole.

\section{Conclusions and future work}
\noindent

In this article, we have studied physical properties of accelerating black holes. These black holes are described by the C-metric, which is not spherically symmetric. This fact motivated us to generalize, for the C-metric, the known relations between eikonal quasinormal modes, the frequencies of circular null geodesics, and the associated Lyapunov exponents, since previous derivations relied on spherical symmetry. Building on this result, we computed the quasinormal modes and greybody factor, in the eikonal limit, for test massless scalar fields in both uncharged and charged accelerating black hole backgrounds. The greybody factors were obtained through their established connection with quasinormal modes. Finally, we obtained an expression for the radius of the shadow cast by an accelerating black hole in terms of the angular velocity of circular null geodesics or the real part of the quasinormal mode frequency, thereby extending another known result from the spherically symmetric case. Setting the acceleration to zero, our results for charged black holes correspond to those of the Reissner-Nordstr\"om solution, which, to the best of our knowledge, has not previously been expressed solely in terms of the black hole parameters.

We have compared our analytical results for the quasinormal mode frequencies to the numerical results of \cite{Destounis:2020pjk}, having obtained an agreement between the two. Since all quantities we have computed can be related to the quasinormal frequencies in this regime, this agreement naturally follows for the greybody factor.

The lack of spherical symmetry of the C-metric also led us to problems such as the introduction of a new separation constant in the scalar field equation of motion. We have obtained the asymptotic behavior of this separation constant in the eikonal limit, and we have shown that such limit is the same for all kinds of perturbations, independently of their spin. From this fact we concluded that the results obtained for quasinormal modes and greybody factor of test massless scalar fields in the eikonal limit can be extended to fields of arbitrary spin.

A full closed form equation for this separation constant is not yet available in the literature; only as a perturbative expansion in the acceleration parameter $a$, supposed to be small. This led us to consider $a$ as a small perturbative parameter throughout the whole article. This approximation is physically reasonable, as one does not expect large accelerations for astrophysical black holes. Yet, by just considering this expansion we are neglecting other possibly interesting regions in parameter space. The WKB method for the quasinormal modes is valid for any parameter region, including the super accelerating regime that we haven't discussed in this article. That may be a topic for future work. We are also planning to study the highly damped limit for the quasinormal modes and greybody factors of this black hole.

\paragraph{Acknowledgements}
\noindent
This work has been supported by Funda\c c\~ao para a Ci\^encia e a Tecnologia under contract IT (UID/50008/2025) and project 2024.04456.CERN. We thank Fundação Calouste Gulbenkian for the program ``New Talents in Physics'', where this work started, and Julián Barragán Amado and Felix Willenborg for fruitful discussions.

\appendix
\section{Generality of the eikonal limit\label{AppendixA}}
\noindent

We now show that the results that we obtained considering massless scalar fields can be generalized for field perturbations of any spin. In order to show this, we need to demonstrate that the eikonal limit of the potential $V_r$ given by \eqref{vreikonal} holds for perturbations of every spin.

This includes showing that the separation constant $\lambda$ does not depend on the spin. In order to reach that conclusion, we need to consider the angular equation and the eikonal limit of the potential $V_\theta$ given by \eqref{vthetaeikonal}. One can use the Newman-Penrose formalism in order to write the Teukolsky equation in the black hole spacetime (\ref{confmetric}). This was done in for a charged black hole with general angular momentum \cite{Wei:2021bqq}. Here we take the angular momentum of the black hole to be zero.

We write the massless field perturbation of spin $p=-s$, as \cite{Wei:2021bqq,Zhou:2024axx,Bini:2008mzd}

\begin{equation}
    \Psi = r^{-s}\Sigma^{s+1}e^{-i\omega t}e^{i m \varphi}R(r) S(\theta)
\end{equation}
where $\Sigma$ is the conformal factor introduced in \eqref{confmetric}. The radial master equation becomes then
\begin{equation}
    R''+\frac{(r^2f(r))'}{r^2f(r)}R'+\left[\frac{\omega^2}{f^2(r)}+\tilde V\right]R=0
\end{equation}
where $B$ is the separation constant and the prime denotes derivative with respect to $r$ and
\begin{equation}
    \tilde V=\frac{(r^2f(r))''}{6r^2f(r)}+s\left( \frac{4i \omega}{rf(r)}-i\frac{\omega (r^2f(r))'}{f(r)}\right)-s^2\left(\frac{((r^2f(r))')^2}{4(r^2f(r))^2}-\frac{(r^2f(r))''}{3r^2f(r)} \right)-\frac{B}{r^2f(r)}
    \label{teukolskyradialpotential}
\end{equation}

We now want to write this equation as in the Schr\"odinger-like form \eqref{waveeq}; we thus introduce
\begin{equation}
    \Phi(r)=\frac{R(r)}{r}\quad \text{and}\quad x=\int\frac{dr}{f(r)}.
\end{equation}
from which we get
\begin{equation}
    \frac{d^2\Phi(r)}{dx^2}+\left( \omega^2-V_r(r)\right)\Phi(r),
    \label{teukolskyradialequation}
\end{equation}
with the potential given by
\begin{equation}
    V_r(r)=-\left(\tilde Vf^2(r) -\frac{ff'}{r}\right).
    \label{teukolskyradialpotential2}
\end{equation}

The equation governing the angular part can be written as
\begin{equation}
    \frac{d^2S}{dz^2}-(m^2-V_\theta)S=0,
    \label{teukolskyangularequation}
\end{equation}
where
\begin{equation}
\begin{aligned}
    V_\theta=-\Big\{s^2\left( -\frac{P(P \sin^2\theta)''}{3}+\frac{P(P\sin^2\theta)' \cos\theta}{3\sin\theta}+\frac{((P\sin^2\theta)')^2}{4 \sin^2\theta}\right)-\\-s\frac{m(P \sin^2\theta)'}{\sin\theta}
    -\frac{P(P\sin^2\theta)''}{6}+\frac{P(P \sin^2\theta)'\cos{\theta}}{6\sin\theta}-PB\sin^2\theta\Big\}.
    \label{teukolskyangularpotential}
\end{aligned}
\end{equation}
where the prime now denotes derivative with respect to $\theta$.

We must relate the separation constants $B$ from \eqref{teukolskyradialpotential} and \eqref{teukolskyangularpotential} to $\lambda$ from  \eqref{waveeq}. Considering $s=0$, we have
\begin{equation}
    V_\theta=P(\theta) \sin^2\theta(B-1/3-5a^2Q^2/3+2Ma\cos \theta +2Q^2a^2\sin^2{\theta})
\end{equation}
which is identical to \eqref{expandvt} if $B=\lambda$.
Since the dependence of the potential on $s$ is of the form of a sum of an $s=0$ term and additive correction $s$, we may assume the same kind of dependence for the separation constant: $B=\lambda+F(s)$, with $F(0)=0$.

In order to obtain the eikonal limit of $V_\theta$ for general $s$, we should take the asymptotic limit $\lambda\rightarrow \infty$. 
In this way, the potential in \eqref{teukolskyangularpotential} becomes
\begin{equation}
    V_\theta=\lambda P(\theta) \sin^2\theta
\end{equation}
which is identical to \eqref{vthetaeikonal2}. Since no other term in \eqref{teukolskyangularequation} depends on $s$, we conclude that, in this limit, all spin perturbations have the same angular potential, and $\lambda$ is also spin independent.\footnote{Notice that $m$ in \eqref{teukolskyangularpotential} is the azimuthal angular quantum number, which obviously does not depend on $s$. The same cannot be immediately concluded about $\omega$ in \eqref{teukolskyradialequation}. The argument that we use here can only be applied to the angular equation.}

We now consider the radial equation \eqref{teukolskyradialequation}. Taking $s=0$ in its potential \eqref{teukolskyradialpotential2} we get
\begin{equation}
    V_r(r)=f(r)\left(\frac{B-1/3-a^2Q^2/3}{r^2}+\frac{2M}{r^3}-\frac{2Q^2}{r^4} \right),
\end{equation}
which becomes identical to $V_r(r)$ obtained in \eqref{expandvr} as long as $B=\lambda$. This is consistent with the previous result obtained with the angular equation.

We now consider the limit $\lambda \rightarrow\infty$ for general $s$. After taking this limit we are left with
\begin{equation}
    V_r=\lambda \frac{f(r)}{r^2},
\end{equation}
which is identical to \eqref{vreikonal} and independent of $s$.
We conclude that the eikonal limit is universal for different kinds of perturbations.


\end{document}